\documentclass[aps,prl,reprint,nofootinbib]{revtex4-1} 

\usepackage{graphicx} 
\usepackage{amsmath,amssymb}
\usepackage[usenames,dvipsnames]{xcolor}
\usepackage{float}
\usepackage{units}
\usepackage{changes}

\usepackage{hyperref}  
\hypersetup{colorlinks = true, citecolor = blue, linkcolor = blue }

\usepackage{soul}

\usepackage{color}
\definecolor{dred}{rgb}{0.80,0.00,0.00} 
\definecolor{dgreen}{rgb}{0.00,0.60,0.00} 
\definecolor{dblue}{rgb}{0.00,0.00,0.80} 
\definecolor{dmagenta}{rgb}{0.80,0.00,0.80}

\begin{document} 

\title{The end of the nucleon-spin crisis}

\newcommand{\mpik}{\affiliation{Max-Planck-Institut f\"ur Kernphysik, 69029 Heidelberg, Germany}}
\newcommand{\kph}{\affiliation{Institut f\"ur Kernphysik, Johannes  
 Gutenberg-Universit\"at Mainz, 55128 Mainz, Germany \\ }}  

\author{Bogdan Povh}\mpik 
\author{Thomas Walcher}\kph

\date{\today}  

\begin{abstract} 
The small fraction of the quark polarization of the nucleon obtained
in deep inelastic lepton scattering is a consequence of a wrong assumption on the 
flux of polarized virtual photons in the analysis of the data. The true flux of polarized 
photons is due to fluctuations into strongly interacting quark-antiquarks at least 20\% 
smaller than assumed and, therefore, the quark polarization is larger. Realizing this 
peculiarity the quark polarization agrees with that given by the Ellis-Jaffe sum rule, 
i.e. is consistent with the prediction derived from the experimentally determined weak 
coupling constants. It also solves the problem with the under-exhaustion of the 
fundamental Bjorken-sum rule. However, it has to be realized that the fluctuations 
render it impossible to measure the quark spin with credibility.
\end{abstract} 

 
\maketitle  

\section{Introduction} \label{intro}
Since the first measurement of the spin polarization of the quarks in the proton in deep 
inelastic scattering (DIS) of muons \cite{Ashman:1989ig} the fraction of the spin  carried by 
the quarks was found to be much smaller than the value  $\Delta\Sigma = 0.59$ expected 
by the Ellis-Jaffe sum rule \cite{Ellis:1973kp} deduced from the SU(3) constituent 
quark model.  This finding was confirmed by the subsequent experiments  at CERN of the 
SMC Collaboration \cite{Adeva:1998vv} and at SLAC by the E143 Collaboration 
\cite{Abe:1998wq} and called somewhat caricaturing the "nucleon-spin crisis". 
A detailed summary is given in  \cite{Aidala:2012mv}. 

The recent measurement of the HERMES Collaboration at HERA 
\cite{Airapetian:2006vy} using the "modified minimal subtraction" (MS) scheme 
\cite{Bardeen:1978yd} determines that only $\Delta\Sigma=0.330 \pm 0.025$ of 
the spin momentum of the nucleon is contributed by the quarks if measured at a 
resolution of Q$^2 = 5$ GeV$^2$. The precision is much improved due to a 
measurement on the deuteron. We use for the presentation in this letter the precise 
measurement of the HERMES Collaboration and the corrections applied by them 
\cite{Airapetian:2006vy}. A more recent publication of the COMPASS collaboration 
\cite{Adolph:2015saz} is somewhat less precise and has problems which will be 
discussed later.  Any further theoretical uncertainties are not included and we refrain 
from listing them since they are not relevant for the further discussion.

The quark polarization of the nucleon obtained from the constituent quark model 
of Cabbibo \cite{Cabibbo:1963yz,Cabibbo:2003yz} is essentially determined by the 
values of the measured axial-vector coupling constant $g^{np}_{\rm A}$ from the 
neutron decay and the axial-vector coupling for the strangeness changing $\Sigma$
decay $g_A^{\Sigma n}$. Together with the Ellis-Jaffe sum rule one obtains 
$\Delta \Sigma = 0.59 \pm 0.03$.

So far there were many attempts to explain the surprising small fraction of the spin 
carried by quarks derived from DIS. In the frame work of SU(3) it is not possible to reconcile  
the two experimental results:  the quark polarization fraction derived from the weak coupling 
constants and the Ellis-Jaffe sum rule, and the fraction obtained from DIS.  The weak 
interaction is  strictly spin dependent and because of the short range of the interaction no 
distortion is expected. On the other hand, the electromagnetic interaction at high lepton 
energies is contaminated by strong interaction and may manifest strong distortion.  This 
aspect has been ignored until today and it seems to be justified to revisit the analysis of the 
measurement of the quark polarization in DIS and ask for its validity.

 \section{Measurement with polarization} \label{meas}
 The polarized structure functions are derived from a measurement of the cross section 
 asymmetry of the scattering of longitudinally polarized leptons on longitudinally polarized nucleons.
 The asymmetry $A$ is determined in subsequent runs with the nucleon spin orientated parallel and
 antiparallel to the polarized virtual photon normalized to the integrated luminosity in dependance of the 
 Bjorken $x$. It is divided by the sum of the cross sections of the two relative spin orientations 
 so that the integrated flux cancels:
 
 \begin{equation}
 A = \frac{d \sigma^{\uparrow \downarrow}-d \sigma^{\uparrow \uparrow}}{d \sigma^{\uparrow \downarrow}
         +d \sigma^{\uparrow \uparrow}}.
 \label{eq:1}
 \end{equation}
  
 The spin structure functions are, however, proportional to the absolute difference of the 
 cross sections of the two relative spin orientations. Therefore,  in the analysis of all experiments 
 the asymmetries were multiplied by the high statistics unpolarized structure function. 
 These structure functions have been measured by assuming that the flux of virtual photons is that 
 presupposed in the QED formulae for the scattering by the electromagnetic interaction.  We argue 
 in this letter that this assumption is wrong.
 
 In order to see this we must realize that DIS is not an entirely perturbative process as has been pointed 
 out already early by several authors \cite{Bjorken:1973gc,Frankfurt:1988nt,Nikolaev:1990yw,Nikolaev:1990ja}. 
 The photon may fluctuate in a quark-antiquark pair representing a hadronic component in the virtual exchange 
 photon. These longitudinal quark-antiquark pairs loose their energy by the emission and absorption of soft 
 gluons. Consequently, the cross section of the longitudinal quark-antiquark pairs cannot  have any noticeable 
 spin dependence and a fraction of the virtual photon flux has to be considered as unpolarized. 
 
 Experimentally the existence of these pairs and their interaction length has been demonstrated in the shadowing of DIS on nuclei in the quark sea at $x < 0.1$  \cite{Kopeliovich:2012kw}. Before we discuss the consequences we want to look in more detail at the fluctuations. This is best done if the DIS is viewed in the nucleon rest frame \cite{Bauer:1977iq}. The predominant part of the quark-antiquark pairs are small with a dimension given by $Q^2$. Their cross section is  
 $\sigma_{tot}^{h\,N} \propto 1/Q^2$ and is spin dependent. But there are also the longitudinal 
 quark-antiquark pairs of hadronic size $\mu$ with a transverse hadronic cross section 
 $\sigma_{tot}^{h\,N} \propto 1/\mu^2$.  In the discussion of this component for the shadowing of photons in 
 nuclei it is assumed that these pairs  contribute to the sea quarks of the structure function. This assumption is justified by the consideration of the so called "Joffe time" $t = 1/(2 m_p x)$ \cite{Kopeliovich:1995ju,Kopeliovich:2012kw}. At small $x < 0.1$, the sea quark region, this length is large $ t > \mu = 1$\,fm.  For larger $x$ this length gets smaller than the hadronic scale suggesting that the assumption of a point probe is correct. However, in the rest frame of the hadronic fluctuation the correct scale is $\mu$ in all coordinates. If one Lorentz transforms this sphere into the rest frame of the proton it stays a sphere of the same scale.  \hspace{-4mm} \footnote{This observation is strange to many who assume that this transformation produces a pancake shape \cite{Einstein:1905ve,Kopeliovich:2012kw}. However, it has been recognized since long that this is not correct \cite{Lampa:1924xx,Terrell:1959zz}. Roger Penrose also presented the general proof that a sphere stays a sphere for all angles of observation \cite{Penrose:1959vz}. It just rotates. "Observation" means here the requirement that all particles mediating the interaction -  here the gluons with the speed of light - from the surface of a sphere arrive at the same time at the point of observation, i.e. of interaction. This has to be distinguished from Einsteins measurement with synchronized clocks  \cite{Taylor:1992}.} This means that one has to assume a  soft non-perturbative component with the typical hadronic scale also at $ 0.1 \leqslant  x \leqslant 1$ in the interaction of the virtual photon with the nucleon. 
 
In \cite{Kopeliovich:1995ju} it was argued that the weight of the small hard pairs is about 1, whereas the probability of the large soft pairs is suppressed by $W_h^{\gamma^{\star}} = \mu^2/Q^2$. Considering the argument on the longitudinal scale of the fluctuation this is valid for all $x$. This means that in the product $W_h^{\gamma^{\star}} \sigma_{tot}^{h\,N}$ the hadronic scale cancels and both transverse cross sections, the hard and the soft fluctuations are of the same order of magnitude.  As has been pointed out in \cite{Kopeliovich:1995ju} not only the perturbative small hard fluctuations,  but also the soft non-perturbative fluctuations show Bjorken scaling.
 
A more quantitative way to estimate the non-perturbative contribution in DIS is by considering the ratio of the integrated diffractive and the total cross sections. Diffractive cross section in DIS means the scattering on a color neutral fluctuation of the nucleon and corresponds to the soft interaction described in the nucleon rest frame. Such measurements have  been performed at HERA by the ZEUS and H1 Collaborations, in particular as semi inclusive measurements of the leading proton and neutron \cite{Chekanov:2009ac,Aaron:2010ab}. In both experiments the DIS takes place on a color singlet fluctuation with a leading proton or neutron proceeding unhindered in forward direction. These events show up in the deep quark sea at small $x < 0.01$ and represent a good estimate of the soft non-perturbative component in the proton. ZEUS gets a ratio $f = 0.36$ \cite{Derrick:1994dt} and H1 gets $f = 0.26 \pm 0.06$ \cite{Ahmed:1992qc} in agreement with the estimate of Halina Abramowicz \cite{Abramowicz:2004dt} $f = 0.30 \dots 0.40$ 
 
 The consequence of this situation is that one has to consider two parts in the flux of the photon beam: one polarized hard part and one non polarized soft part with a relative interaction probability of $r = 1-f$ ($0 \leqslant r \leqslant 1$) and $f$ respectively. However, it is evident that the polarized cross sections have to be normalized to the polarized and not to the total flux. There are two ways to look at the correction for these two components (see the formulae in e.g. \cite{Ashman:1989ig}):
 \begin{itemize} 
 \item In the asymmetry in Eq.\,(\ref{eq:1}) a correction factor to the flux cancels, but multiplying with the unpolarized structure function means effectively to normalize to the total flux. Therefore, the flux is a factor of $1/r$ to large and the structure function has to be divided by $r$.
 \item The unpolarized structure function appears in the denominator of the asymmetry multiplied with the structure function and  a correction cancels again. One is then left with the polarized cross sections wrongly normalized to the total flux and one has to multiply the absolute difference of the polarized structure functions by $1/r$.
 \end{itemize}
 This means that one has to multiply all results of the polarized structure functions and the integrals derived from them by $1/r$.  
 
\begin{figure}
\includegraphics[width=0.9\columnwidth]{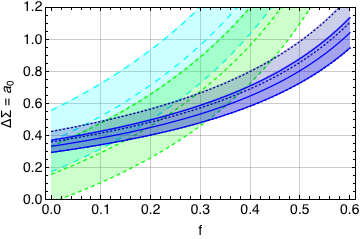}
\caption{The spin fraction $\Delta \Sigma = a_0$ as a function of the fraction of the soft non-perturbative hadronic fraction $f = 1-r$ in the photon beam. The error bands include all errors of $\Gamma_1$ and of the weak axial charge $a_3$ and the weak hyper charge $a_8$. Solid line (blue): HERMES deuteron \cite{Airapetian:2006vy}, short dashed (dark blue): COMPASS deuteron \cite{Adolph:2015saz}, medium dashed (green: HERMES proton  \cite{Airapetian:2006vy}, long dashed (light blue): COMPASS proton \cite{Adolph:2015saz}. (color online) }
\label{fig:a0}
\end{figure}

 The spin fraction carried by the quarks $\Delta \Sigma$ is identical to the singlet axial coupling $a_0 =  \Delta \Sigma$ and calculated from the integral over the the polarized deuteron structure function $\Gamma_1^d$ \cite{Airapetian:2006vy}:
\begin{equation}
 a_0 (Q_0^2) =  \frac{1}{\Delta C_S^{\overline{MS}}} \left[ \frac{9\, \Gamma_1^d / r}{(1-\frac{3}{2} \omega_D)} - 
 \frac{1}{4} a_8 \Delta C^{\overline{MS}}_{NS} \right] 
 \end{equation}
 where the $\Delta C_S^{\overline{MS}}$ and $\Delta C_{NS}^{\overline{MS}}$ are the singlet and non-singlet Wilson coefficient functions with $Q_0^2 = 5\,$GeV$^2$ ($\alpha_s = 0.29 \pm 0.01$) und $\omega_D = 0.05 \pm 0.01$ is a correction for the binding of the deuteron. For the proton the equivalent relation
\begin{equation}
a_0 (Q_0^2) =  \frac{1}{\Delta C_S^{\overline{MS}}} \left[ 9\, \Gamma_1^p / r - \frac{3}{4} a_3 \Delta C^{\overline{MS}}_{NS}- 
 \frac{1}{4} a_8 \Delta C^{\overline{MS}}_{NS} \right] 
  \end{equation}
holds where we assume the most recent values \cite{Eidelman:2004wy} for the weak charges $a_3 =  g_{\rm A} = 1.269 \pm 0.003$ and $a_8 = 0.586 \pm 0.031$. The assumption of this value for $a_3$ given by the neutron beta decay means the exact exhaustion of the Bjorken-sum rule. For further details we refer to the paper of the HERMES Collaboration \cite{Airapetian:2006vy}. The dependence of $\Delta \Sigma$ represented by these equations as a function of $f= 1-r$ is depicted in Fig.\,\ref{fig:a0} 
The Ellis-Jaffe-sum rule requires that $a_0 = \Delta \Sigma= 0.59$. For the HERMES deuteron one reads $f_d = 0.36 \pm 0.05$ and for the COMPASS deuteron the consistent value $f_d = 0.32 \pm 0.08$, for the HERMES proton $f_p = 0.27 \pm 0.05$ and for the COMPASS proton the less consistent value $a_0 = 0. 15\pm 0.10$. However, for the deuteron consisting of a bound proton and neutron with a size of the strong scale one has to realize that the absorption of a soft non-perturbative fluctuation of the same scale will be coherent. Therefore $f_d$ will be larger than $f_p$ since the cross section of the deuteron for the fluctuation will be larger.  This is in contrast to the correct assumption of incoherent scattering in DIS, i.e. the linear addition of the photon cross sections. 

A remark is in place on the difference of the $f_p^{\textrm{HERMES}}$ and $f_p^{\textrm{COMPASS}}$. The HERMES proton data are based on their own unscaled measurement and due to its insufficient accuracy they do not extrapolate to $x \rightarrow 0$ and refrain from comparing to the Bjorken-sum rule  \cite{Airapetian:2006vy}.  COMPASS  uses a fit of the "world data" of $\Gamma_1^p$ and determines $a_3$ from $\Gamma_1^{\textrm{NS}}$, i.e. checks the Bjorken-sum rule, however without looking at the spin content \cite{Adolph:2015saz}. In their fits they extrapolate and recalibrate the data of other measurements - the one of HERMES by as much as 10\% - and still their own data are below these fits. It is understandable that they try to fathom the possibilities to proof the fundamental Bjorken-sum rule from the existing data, however with the realization of the idea presented in this paper the Bjorken sum as well as the Ellis-Jaffe-sum rule are in accord with the data naturally.
 
 \section{Discussion} \label{disc}
In the following we want to discuss the consequences for the quark distributions if one assumes the natural values for the weak coupling constants $a_0 = \Delta \Sigma = 0.59 \pm 0.1$, $a_3$, and $a_8$ as given above. The large error of $a_0$ is an ad-hoc assumption of our lack of knowledge of the fraction of the soft fluctuations and their cross sections.
The quark distributions for the lightest three flavors $u,d,s$  (see e.g. \cite{Airapetian:2006vy}) can be expressed as the singlet charge $a_0$, the axial vector charge $a_3 = g_{\rm A}$ and the hyper charge $a_8$. In the SU(3) constituent quark model the quark helicity distributions read:

\begin{align}
a_0 = & (\Delta u + \Delta \bar{u}) +  (\Delta d + \Delta \bar{d}) +  (\Delta s + \Delta \bar{s}) = \Delta \Sigma \\
a_3 = & (\Delta u + \Delta \bar{u}) -  (\Delta d + \Delta \bar{d}) \\
a_8 = & (\Delta u + \Delta \bar{u}) +  (\Delta d + \Delta \bar{d}) - 2 (\Delta s + \Delta \bar{s})
\end{align}
 
With the correct value for $a_0 = 0.59 \pm 0.1$ one gets :
 
 \begin{alignat}{2}
 (\Delta u + \Delta \bar{u}) &=   & 0.932 \pm 0.034 \\
 (\Delta d + \Delta \bar{d}) &= - & 0.337 \pm 0.034 \\
 (\Delta s + \Delta \bar{s}) &=  & 0.005 \pm 0.034
 \end{alignat}
 
This has to be compared to the values with the uncorrected $a_0 = 0.330 \pm 0.025$ of HERMES  \cite{Airapetian:2006vy}.
 
 \begin{alignat}{2}
 (\Delta u + \Delta \bar{u}) &=   & 0.843 \pm 0.014 \\
 (\Delta d + \Delta \bar{d}) &= - & 0.427 \pm 0.014 \\
 (\Delta s + \Delta \bar{s}) &=  - & 0.087 \pm 0.017
 \end{alignat}
 
The large errors of the corrected values contain the range of the fraction of the soft non-perturbative component $f$ not existing for the assumption of a 100\% polarized photon beam. 
 
\begin{figure}
\includegraphics[width=0.95\columnwidth]{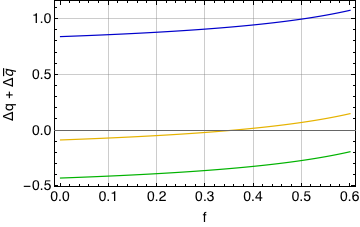}
\caption{The quark distributions $\Delta q + \Delta \bar{q}$ for the flavors u (upper), s (middle), d (lower) as a function of the fraction of the soft non-perturbative hadronic fraction $f = 1-r$ in the photon beam.  (color online)}
\label{fig:uds}
\end{figure}

 In Fig.\,\ref{fig:uds} we show the quark distributions again as a function of the hadronic fraction $f$.
Particularly the reduction of the contribution of the strange sea is remarkable, now compatible with no contribution of the strange sea to the polarization meaning that the SU(2) description would be sufficient or at least a good approximation. Of course, this simple calculations neglect any next-to-leading order corrections or finite quark masses breaking the SU(3) symmetry. But the values give an indication how dramatic our idea about the spin composition changes with the correct value for $a_0 = \Delta \Sigma$. The small or vanishing strange quark contribution remedies also the difference of the strange quark-helicity distributions of the HERMES result given above from earlier HERMES results obtained by semi-inclusive DIS \cite{Airapetian:2006vy}.

It is not very likely that one can improve the accuracy of the ratio of hard photon and soft non-perturbative hadron parts theoretically. The measurement of the spin polarization in the deep inelastic scattering will thus not improve our knowledge of the quark spins beyond the information from the neutron and hyperon decays. However, the importance of the non-perturbative effects in the deep inelastic scattering became very clear. 

It is interesting to note that the standard evolution of the structure function does not consider the non-perturbative effects. 
Going back to the system in which the nucleon is in the infinite momentum frame and realizing that  about 30\% of the events go via the non-perturbative scattering it becomes clear how naive is the picture of the noninteracting partons of the nucleon in the Weizs\"acker-Williams system. The shape of the structure function is controlled with roughly equal weight  by the perturbative as well as non-perturbative processes. In spite of this observation the structure functions can be forced in the perturbative corset since both the perturbative and the non-perturbative deep inelastic scattering obey Bjorken scaling \cite{Kopeliovich:1995ju}.

\section{Conclusion} \label{concl}
In conclusion we have presented a solution for the much discussed "spin crisis" of the nucleon: the valence quarks carry the major part of the nucleon spin indeed. This solution emerges naturally from the realization that the photon probe contains a hard part, the electromagnetic interaction proper, and a soft non-perturbative part due the quark-antiquark fluctuations.  The old natural idea that the nucleon can be well described as composed of constituent massive quarks which was so successful in explaining the excitation spectrum of the nucleon, the baryon spectrum, and the ratio of the magnetic moments is reestablished. The experimental finding that the nucleon is a strongly coupled relativistic many body system which shows in leading order scaling of point like almost massless quarks is not in contradiction since the soft non-perturbative hadronic part of the photon also scales as argued. Depending on the resolution of the probe one finds scaling violations which converge  in the limit of vanishing momentum transfer to the effective constituent quarks. We have shown that this idea can be maintained. 

\begin{acknowledgments}
We are indebted to Boris Kopeliovich  and Mitja Rosina for many clarifying discussions of the physics of this paper over years, and Jan-Erik Olsson  for providing explanations and details of the H1 and ZEUS results.
\end{acknowledgments}

\bibliography{references}

\end{document}